\documentstyle[12pt]{article}
\begin{document}

\def\p{\phi}
\def\P{\Phi}
\def\a{\alpha}
\def\e{\varepsilon}
\def\be{\begin{equation}}
\def\ee{\end{equation}}
\def\l{\label}
\def\0{\setcounter{equation}{0}}
\def\T{\hat{T}_}
\def\b{\beta}
\def\S{\Sigma}
\def\3{d^3{\rm \bf x}}
\def\4{d^4}
\def\C{\cite}
\def\r{\ref}
\def\ba{\begin{eqnarray}}
\def\ea{\end{eqnarray}}
\def\n{\nonumber}
\def\R{\right}
\def\L{\left}
\def\q{\hat{Q}_0}
\def\X{\Xi}
\def\x{\xi}
\def\la{\lambda}
\def\d{\delta}
\def\s{\sigma}
\def\f{\frac}
\def\vx{{\rm \bf x}}
\def\j{\frac{\delta}{i \delta j_a ({\rm \bf x},x_0+t+t_1)}}

\begin{titlepage}
\begin{flushright}
      {\normalsize IP GAS-HE-5/95}
\end{flushright}
\vskip 3cm
\begin{center}
{\Large\bf $S$-matrix interpretation of  finite-temperature
real-time field theories}
\vskip 1cm

\mbox{J.Manjavidze}\footnote{Institute of Physics,
Georgian Academy of Sciences, Tamarashvili str. 6,
Tbilisi 380077, Republic of Georgia,
e-mail:~jm@physics.iberiapac.ge} \\

\end{center}
\date{MARCH 1995}
\vskip 1.5cm

\begin{abstract}
\footnotesize
    The aim of  the article is to construct the $S$-matrix
interpretation of the perturbation theory for the Wigner functions
generating functional at a finite temperature. The temperature is
introduced in the theory by the way typical for the microcanonical
description. The perturbation theory contains the two-temperature
Green functions. The two possible boundary conditions are considered.
One of them is usual in a field theory vacuum boundary condition.
The corresponding generating functional can be used in the particle
physics. Another type of the boundary condition assumes that the system
under consideration is in environment of the free particles background
field. This leads to the theory with Kubo-Martin-Schwinger boundary
conditions ones at the one-temperature limit.

\end{abstract}

\end{titlepage}

\section{Introduction}
\setcounter{equation}{0}

The field-theoretical description of statistical systems at a finite
temperature is usually based on the formal  analogy between imaginary
time and inverse  temperature $\b$ ($\b=1/T$) \C {bl}. This approach
is fruitful \C {mats} for description of the static properties of a system,
but it demands a complicated mathematical apparatus for the
analytic continuation to the real time \C {land}, if we want
to clear up the dynamical aspects. At the  same time a great number
of modern problems is  connected with system dynamics. It is
important, for instance, for description of evolution of the
Universe, of quark-gluon plasma behavior, etc.

The  first important quantitative attempt to build the real-time
finite-temperature field theory \C {jack} discover a problem of the
pinch-singularities. The further investigation of the theory have
allowed to demonstrate the  cancellation mechanism of  these unphysical
singularities \C {sem}. This attained by doubling of the degrees of
freedom \C {sch,kel}:  the Green functions of the theory represent $2$
$\times$ $2$ matrix.  It surely makes the theory more complicated, but
the operator formalism of the  thermo-field dynamics \C {um} shows the
unavoidable character  of this complication.

It is  important to note that the traditional real-time finite-temperature
field-theoretical description \C {sch,kel} of statistical systems based on the
Kubo-Martin-Schwinger (KMS) \C {sch,mar,kubo} condition for a field:
\be
\Phi (t)=\Phi (t-i\beta)
\l{1}
\ee
which, without fail, leads to the $equilibrium$ fluctuation-dissipation
conditions \C {haag} (see also \C {chu}).

We shell use the $S$-matrix approach which is natural for
description of time evolution. (The $S$-matrix  description was used also
in \C {dash,carr}.) For this purpose the amplitudes
\be
<(p)_m|(q)_n>=a_{n,m}(p_1,p_2,...,p_m;q_1,q_2,...,q_n)
\l{2}
\ee
of the $n$- into $m$-particles transition will be introduced (Sec.2). The
in- and out-states must be composed from mass-shell particles \C {pei}.
Using these amplitudes we will calculate the probability which is
\be
\sim |a_{n,m}|^2=<(p)_m |(q)_n ><(q)_n |(p)_m >.
\l{3}
\ee
This will lead to the doubling of degrees of freedom.

The  finite temperature  will be introduced (see also \C {kaj}) taking into
account that, for instance,
\be
d \Gamma_n =|a_{n,m}|^2 \prod_{1}^{n}\frac{d^3 q_i}{(2\pi )^3 2\epsilon
(q_i)} , \;\;\; \epsilon (q)= (q^2 +m^2)^{1/2},
\l{4}
\ee
is  the differential measure  of  the initial state. Then we will
define the temperature as the function of initial energy through
the equation of state (see Sec.2). This introduction of
temperatures as the Lagrange  multiplier  is  obvious for
microcanonical description \C {mar}.

The final-state temperature will be introduced by the same way.
So, we will construct the two-temperature theory.  In this
theory with  two temperatures (the equation of state can be applied
at  the very end of calculations) it is impossible to use the KMS
boundary condition.

One can note that  the product of  amplitudes  (\ref{3}) describes the
closed-path motion in the functional space, with some
``turning-point" fields $\P (\s_{\infty})$, where $\s_{\infty}$ is the
infinitely far hypersurface (Sec.3). The value of $\P (\s_{\infty})$
specify the evironment of a system.

In Sec.2  the vacuum boundary condition $\P(\s_{\infty})=0$, familiar for
a field  theory, will be considered. This theory can be applied
in the particle physics. The  simplest (minimal) choice of
$\P (\s_{\infty})\neq 0$ assumes that the  system under consideration
is surrounded by black-body radiation (Sec.4). This interpretation
restores Niemi-Semenoff's formulation of the theory \C {sem}.

One should admit also that this choice of boundary condition is not
unique: one can consider another organization of the environment of
considered system. The $S$-matrix interpretation will be used since it
able to show the way of adoption of the formalism to the arbitrary boundary
conditions. It should broaden the potentialities of the real-time
finite-temperature field-theoretical methods. The special interest
represent the topological effects, but in this paper consideration will be
performed in the perturbation framework only.

\section {Vacuum boundary condition}
\setcounter{equation}{0}

The starting point of our calculations is $n$- into $m$-particles
transition amplitude $a_{n,m}$, the derivation of which is well
known procedure  in the perturbation framework. Let us introduce
$(n+m)$-point Green function $G_{n,m}$ through a generating functional
$Z_{j}$ \C {vas}:
\be
G_{n,m}((x)_n;(y)_m)=(-i)^{n+m}\prod_{k=1}^{n}\hat{j}(x_k)\prod_{k=1}^{m}
\hat{j}(y_k)Z_j,
\l{5}
\ee
where
\be
\hat{j}(x)=\frac{\delta}{\delta j(x)},
\l{*}
\ee
and
\be
Z_j=\int D\Phi e^{iS_j(\Phi)}.
\l{6}
\ee
The action
\be
S_j (\Phi)=S(\Phi)-V(\Phi)+\int dxj(x)\Phi (x),
\l{7}
\ee
where $S(\P)$ is the free part and $V(\P)$ describes the interactions.
At the end one can put $j=0$.

To provide the convergence of the integral (\ref {6}) over scalar field
$\P$ the action $S_{j}(\P)$ must contain positive imaginary part.  Usually
for this purpose Feynman's $i\e$-prescription is used.  It is better for
us to shift infinitesimally the time contour to the upper half plane
\C{mil,land}, i.e. on the following contour
$C_+ :t\rightarrow t+i\epsilon, \;\;\; \epsilon >0$
and after all calculations to
return the time contour  on the real axis, $\epsilon \rightarrow +0$.

In eq. (\ref{6}) the integration is performed over all field
configurations with standard vacuum boundary condition:
\be
\int d^4 x \partial_{\mu}(\Phi \partial^{\mu}\Phi)=
\int_{\sigma_{\infty}}d\sigma_{\mu}\Phi\partial^{\mu}\Phi=0,
\l{8}
\ee
which prosides to zero contribution from the surface term.

Let us introduce now field $\p$ through the equation:
\be
-\frac{\delta S(\phi)}{\delta\phi(x)} =j(x)
\l{9}
\ee
and perform the shift $\P\rightarrow\P+\p$ in integral (\ref{6}),
conserving boundary condition (\ref{8}). Considering $\p$ as the probe
field created by the  source:
\ba
\phi(x)=\int dy G_0 (x-y)j(y),
\n \\
(\partial^2 +m^2)_x G_0 (x-y)=\delta (x-y),
\l{10}
\ea
the only connected Green function $G^{c}_{n,m}$ will be
interesting for us. Therefore,
\be
G_{n,m}^{c}((x)_n;(y)_m)=(-i)^{n+m}\prod_{k=1}^{n}\hat{j}(x_k)
\prod_{k=1}^{m}\hat{j}(y_k)Z(\phi),
\l{11}
\ee
where
\be
Z(\phi)=\int D\Phi e^{iS(\Phi)-iV(\Phi+\phi)}
\l{12}
\ee
is the new generating functional.

To calculate the nontrivial elements of $S$-matrix we
must put the external particles on the mass shell. Formally this
procedure means amputation of the external legs of $G^{c}_{n,m}$
and further multiplication on the free particles wave functions.
In result the amplitude of $n$- into $m$-particles transition $a_{n,m}$
in the momentum representation has the form:
\be
a_{n,m}((q)_n;(p)_m)=(-i)^{n+m}\prod_{k=1}^{n}\hat{\phi}(q_k)
\prod_{k=1}^{m}\hat{\phi}^* (p_k) Z(\phi).
\l{13}
\ee
Here we introduce the operator
\be
\hat{\phi}(q)=\int dx e^{-iqx} \hat{\phi}(x),\;\;\;
\hat{\phi}(x)=\frac{\delta}{\delta \phi (x)}
\l{14}
\ee
and $q_{k}$ and $p_{k}$ are the momentum of in- and out-going particles.

Supposing that the momentum of  particles are insufficient for us the
probability of $n$- into $m$-particles transition is defined by the
integral:
\be
r_{n,m}=\frac{1}{n!m!}\int d\omega_n (q) d\omega_m (p)
\delta^{(4)}(\sum_{k=1}^{n}q_k - \sum_{k=1}^{m}p_k) |a_{n,m}|^2,
\l{15}
\ee
where
\be
d\omega_n(q)=\prod_{k=1}^{n}d\omega(q_k)=
\prod_{k=1}^{n}\frac{d^3  q_k}{(2\pi)^3 2\epsilon (q_k)}, \;\;\;\;
\epsilon =(q^2+m^2)^{1/2},
\l{16}
\ee
is  the Lorentz-invariant phase space element. We assume that the
energy-momentum conservation $\d$-function was extracted from the
amplitude.

Note that $r_{n,m}$ is the divergent quantity. To avoid this problem
with trivial divergence let us divide the energy-momentum fixing
$\d$-function into two parts:
\be
\delta^{(4)}(\sum q_k - \sum p_k)=\int d^4 P
\delta^{(4)}(P-\sum q_k)\delta^{(4)}(P-\sum p_k)
\l{17}
\ee
and consider a new quantity:
\be
r(P)=\sum_{n,m}\frac{1}{n!m!}\int d\omega_n (q) d\omega_m (p)
\delta^{(4)}(P-\sum_{k=1}^{n} q_k)
\delta^{(4)}(P-\sum_{k=1}^{n} p_k) |a_{n,m}|^2.
\l{18}
\ee
Here we suppose that the number of particles are  not fixed. It is
not too hard to see that, up to phase space volume,
\be
r=\int d^4P\; r(P)
\l{19}
\ee
is the imaginary part of amplitude $<vac|vac>$. Therefore, computing
$r(P)$ the  standard renormalization procedure  can be  applied and the
new divergences will not arise in our formalism.

The Fourier  transformation of $\d$-functions in (\ref{18}) allows to
write $r(P)$ in the form:
\be
r(P)=\int \frac{d^4 \alpha_1}{(2\pi)^4}\frac{d^4\alpha_2}{(2\pi)^4}
e^{iP(\alpha_1+\alpha_2)}
R(\alpha_1, \alpha_2),
\l{20}
\ee
where
\be
R(\alpha_1, \alpha_2)=
\sum_{n,m} \frac{1}{n!m!}\int
\prod_{k=1}^{n}\{d\omega(q_k)e^{-i\alpha_1 q_k}\}
\prod_{k=1}^{m}\{d\omega(p_k)e^{-i\alpha_2 p_k}\} |a_{n,m}|^2.
\l{21}
\ee
The introduction of the ``Fourier-transformed" probability $R(\a_1 ,\a_2)$
means only that the  phase-space volume is not fixed exactly, i.e.  it is
proposed that 4-vector $P$ is fixed with some accuracy if $\a_i$ are
fixed. The energy and momentum in our approach are still locally conserved
quantities since an amplitude $a_{nm}$ is translational invariant.  So, we
can perform the transformation:
\be
\a_1 \sum q_k =(\a_1 -\sigma_1 )\sum q_k +\sigma_1 \sum q_k
\rightarrow (\a_1 -\sigma_1 )\sum q_k +\sigma_1 P
\l{}
\ee
since 4-momenta are conserved. The choice of $\sigma_1$ fixes the
reference  frame. This  degree of freedom of the theory was considered
in \C{mta,psf}.

Inserting (\ref{13}) into (\ref{21}) we find  that
\ba
R(\alpha_1, \alpha_2)=\exp \{i\int dx dx'(
\hat{\phi}_+(x)D_{+-}(x-x',\alpha_2)\hat{\phi}_-(x')-
\n \\
\hat{\phi}_-(x)D_{-+}(x-x',\alpha_1)\hat{\phi}_+(x'))\}
Z(\phi_+)Z^* (\phi_-),
\l{22}
\ea
where $D_{+-}$ and $D_{-+}$ are  the positive and negative frequency
correlation functions correspondingly:
\be
D_{+-}(x-x',\alpha)=-i\int d\omega(q)e^{iq(x-x'-\alpha)}
\l{23}
\ee
describes the process of particles creation at the  time  moment $x_0$
and its absorption at $x'_0$, $x_0>x'_0$, and $\a$ is the
center of mass (CM) 4-coordinate. Function
\be
D_{-+}(x-x',\alpha)=i\int d\omega(q)e^{-iq(x-x'+\alpha)}
\l{24}
\ee
describes the opposite process, $x_0<x'_0$. These functions obey the
homogeneous equations:
\be
(\partial^2 +m^2)_x G_{+-}=
(\partial^2 +m^2)_x G_{-+}=0
\l{25}
\ee
since the propagation of mass-shell particles is described.

We suppose that $Z(\p)$ may be computed perturbatively. For this purpose
the following transformation will be used:
\ba
e^{-iV(\phi)}=
e^{-i\int dx \hat{j}(x)\hat{\phi}'(x)}
e^{i\int dx j(x)\phi (x)}
e^{-iV(\phi ')}=
\n\\
e^{\int dx \phi(x)\hat{\phi}'(x)}
e^{-iV(\phi ')}=
\n\\
e^{-iV(-i\hat{j})}
e^{i\int dx j(x)\phi (x)},
\l{26}
\ea
where $\hat j$ was defined in (\ref{*}) and $\hat{\phi}$ in
(\ref{14}). At the end of calculations the auxiliary variables
$j$, $\p'$ can be taken equal to zero. Using the
first equality in (\ref{26}) we find that
\be
Z(\phi)=
e^{-i\int dx \hat{j}(x)\hat{\Phi}(x)}
e^{-iV(\Phi+\phi)}
e^{-\frac{i}{2}\int dx dx'
 j(x)D_{++}(x-x')j(x')},
\l{27}
\ee
where $D_{++}$ is the causal Green function:
\be
(\partial^2 +m^2)_x G_{++} (x-y)=\delta (x-y)
\l{28}
\ee
Inserting (\ref{27}) into (\ref{22}) after simple manipulations with
differential operators, see (\ref{26}) we find the expression:
\ba
R(\alpha_1, \alpha_2)=
e^{-iV(-i\hat{j}_+)+iV(-i\hat{j}_-)}\times
\n \\ \times
\exp\{ \frac{i}{2} \int dx dx'(
 j_+ (x)D_{+-}(x-x',\alpha_1)j_- (x')-
 j_- (x)D_{-+}(x-x',\alpha_2)j_+ (x')-
\n \\
 -j_+ (x)D_{++}(x-x')j_+ (x')+
 j_- (x)D_{--}(x-x')j_- (x'))\},
\l{29}
\ea
where
\be
D_{--}=(D_{++})^*
\l{30}
\ee
is the anticausal Green function. One can consider
$R(\alpha_1,\alpha_2)=R(\alpha_1,\alpha_2;j_1,j_2)$ as the generating
functional for Wigner functions (see also \C {land,hu}).

Considering the system with large number of particles we
can simplify calculations choosing the CM frame $P=(P_0 =E,\vec 0)$.
It is useful also \C {kaj,mar} to rotate the contours of integration
over $\alpha_{0,k}$: $\alpha_{0,k}=-i\b_k, Im\b_k =0, k=1,2$.
In result, omitting unnecessary constant, we will consider
$R=R(\b_1,\b_2)$.

External particles play the double role in the $S$-matrix approach:
their interactions create and annihilate the system under consideration
and, on the other hand, they are probes through which the
measurement of a system is  performed. Since $\b_k$ are the conjugate
to the particles  energies quantities we will interpret them
as  the inverse temperatures in the initial ($\b_1$) and  final ($\b_2$)
states of interacting fields. But there is the question: are constants
$\b_k$ really the  ``good" parameters to describe the system.

The integrals over $\b_k$:
\be
r(E)=\int \frac{d\beta_1}{2\pi i}\frac{d\beta_2}{2\pi i}
e^{(\beta_1 +\beta_2)E}
e^{-F(\beta_1,\beta_2)},
\l{31}
\ee
where
\be
F(\beta_1,\beta_2)=-\ln R(\beta_1,\beta_2),
\l{32}
\ee
can be computed by the stationary phase method. This assumes that the total
energy $E$ is a fixed quantity. The solutions of the equations (of state):
\be
E=\frac{\partial F(\beta_1,\beta_2)}{\partial \beta_k} ,\;\;\;   k=1,2,
\l{33}
\ee
gives the mostly probable values of $\b_k$ at a given $E$. Eqs. (\ref{33})
always  have the real solutions and, because of energy conservation law,
both eqs. (\ref{33}) have the same solution with the  property \C {mar}:
\be
\beta_k=\beta (E), \;\;\;\;    \beta>0.
\l{34}
\ee
If the fluctuations of $\b_k$ are large  it is insufficient to know
$\b(E)$: the expansion of integral (\ref{31}) over $(\b-\b_k)$ will lead
to the asymptotic series with zero convergence radius since
$F(\b_1,\b_2)$ is essentially nonlinear function. In this paper we
assume that $\b$ is the ``good" parameter, i.e. the  fluctuations  of
$\b_k$ are Gaussian. In this case we can interpret $F(\b_1,\b_2)$
as  the free energy and $1/\b_k$ as the temperatures. Such definition
of thermodynamical parameters is in a spirit of microcanonical
description. Note the important role of decomposition (2.14) in this
interpretation of the $\beta _k$.

The structure of generating functional (\ref{29}) is the same as the
generating functional of Niemi-Semenoff \C {sem} have. The difference is
only in the definition of Green functions which follows from the choice
of boundary condition (\ref{8}). The Green functions  $D_{ij}, i,j=+,-$
were defined on the time contours $C_{\pm}$ in the complex time
plane ($C_-=C_+^*$). This definition of the  time  contours coincide with
Keldysh' time contour \C {kel}. The expression (\ref{29}) can be written
in the compact form if the matrix notations are  used. Note also a
doubling of the  degrees of freedom. This doubling is unavoidable
since Green functions $D_{ij}$ are  singular  on the light cone.

\section{Closed-path boundary conditions}
\setcounter{equation}{0}

The calculation of generating functional $R(\a_1,\a_2)$ is performed
introducing the corresponding generating functional
\ba
R_0 (\phi_{\pm})=Z(\phi_+)Z^* (\phi_-)=
\n \\
\int D\Phi_+ D\Phi_-
\exp\{iS(\Phi_+)-iS(\Phi_-)-iV(\Phi_+ +\phi_+) + iV(\Phi_- +\phi_-)\},
\l{35}
\ea
see (\ref{22}). The fields $\p_+,\p_-$ and $\P_+,\P_-$ were defined on the
time contours $C_+,C_-$. By definition, path integral (\ref{35}) describes
the closed path motion in the space of fields $\P$. We want to use this fact
and introduce a more general boundary condition which also guaranties the
cancelation of the surface terms in the perturbation  framework. We
will introduce the equality:
\be
\int_{\sigma_{\infty}} d\sigma_{\mu} \Phi_+  \partial^{\mu}\Phi_+ =
\int_{\sigma_{\infty}} d\sigma_{\mu} \Phi_-  \partial^{\mu}\Phi_-.
\l{36}
\ee
The solution of eq.(\ref{36}) requires that the fields $\P_+$ and $\P_-$
(and theirs first derivatives $\partial_{\mu}\P_{\pm}$) coincide on the
boundary hypersurface $\s_{\infty}$:
\be
\Phi_{\pm}(\sigma_{\infty})=\Phi(\sigma_{\infty}),
\l{37}
\ee
where,  by definition, $\Phi(\sigma_{\infty})$
is the arbitrary, ``turning-point", field.

In absence of the surface terms, the existence of nontrivial field
$\Phi(\sigma_{\infty})$ has  the influence only on the  structure of Green
functions
\ba
G_{++}=<T\Phi_+\Phi_+>,\;\;\;
G_{+-}=<\Phi_+\Phi_->,
\n \\
G_{-+}=<\Phi_-\Phi_+>, \;\;\;
G_{--}=<\tilde{T}\Phi_-\Phi_->
\l{38}
\ea
where $\tilde{T}$ is the untitemporal time ordering operator. This Green
functions must obey the  equations:
\ba
(\partial^2 +m^2)_x G_{+-} (x-y)=
(\partial^2 +m^2)_x G_{-+} (x-y)=0,
\n \\
(\partial^2 +m^2)_x G_{++} (x-y)=
(\partial^2 +m^2)_x^* G_{--} (x-y)=\delta (x-y),
\l{39}
\ea
and the general solution of these equations:
\ba
G_{ii}=D_{ii}+g_{ii},
\n \\
G_{ij}=g_{ij},\;\;\;\; i\neq j
\l{40}
\ea
contain the undefined terms $g_{ij}$ which are the solutions of
homogenous equations:
\be
(\partial^2 +m^2)_x g_{ij} (x-y)=0,\;\;\; i,j=+,-.
\l{41}
\ee
The general solution of these equations (they are distinguished by the
choice of the time contours $C_{\pm}$)
\be
g_{ij}(x-x')=\int d\omega (q) e^{iq(x-x')} n_{ij} (q)
\l{42}
\ee
are defined through the functions $n_{ij}$ which are the
functionals of ``turning-point" field $\Phi(\sigma_{\infty})$: if
$\Phi(\sigma_{\infty})=0$ we must have $n_{ij}=0$ and we will come back
to the theory of previous section.

Our aim is to define $n_{ij}$. We can suppose that
$n_{ij}\sim <\Phi(\sigma_{\infty})\cdots\Phi(\sigma_{\infty})>$.
The simplest supposition gives:
\be
n_{ij}\sim <\Phi_{i}\Phi_{j}>\sim <\Phi^2(\sigma_{\infty})>.
\l{43}
\ee
We will find the exact definition of
$n_{ij}$ starting from the $S$-matrix interpretation of the  theory.

\section{KMS boundary condition}
\setcounter{equation}{0}

In the previous section it was shown that the theory permits the
arbitrariness of boundary condition: the turning-point field
$\P(\s_{\infty})$ may be arbitrary since the ``closed-path" motion
in the functional space is described. We will suppose that on the
infinitely far hypersurface $\s_{\infty}$ there are only free,
mass-shell, particles. Formally it follows from (\ref{40}) -(\ref{42}).
This assumption is natural also in the $S$-matrix framework \C{pei}.
In other respects the choice of boundary condition is arbitrary.

Therefore, our aim is  concerned with description of evolution of the
system in a background field of mass-shell particles. In this paper
we will assume that there are not any special correlations among
background particles. We will take into account only the restrictions
connected with energy-momentum conservation laws. Quantitatively this
means that multiplicity distribution of background particles is
Poison-like, i.e. is  determined by the mean multiplicity only.
This is in spirit of definition of $n_{ij}$ in eqs. (\ref{42}),
(\ref{43}).

Our derivation is the same as in \C {psf}. Here we restrict ourselves
mentioning only the main quantitative points.

Calculating the  product $a_{n,m}a^*_{n,m}$ we describe a process of
particles creation and further their adsorption. In the vacuum case
of Sec.2 the two process were taken into account: creation of
particles by the $a_{n,m}$ and theirs adsorption by $a^*_{n,m}$ ,
and the opposite process of particles creation by $a^*_{n,m}$ and
theirs adsorption by $a_{n,m}$. This  processes were time ordered.
This was the reason of the frequency correlation functions $D_{+-}$
and  $D_{-+}$ appearance. In the nonvacuum case, i.e. in presence  of
the background particles, this  time-ordered picture is slurring over
since the possibility to absorb particles before their creation appears.

The processes of creation and adsorption are described in vacuum
by the product of operator exponents of $\hat\p_+\hat\p_-$,
$\hat\p_-\hat\p_+$. We can derive (see also \C {psf}) the generalizations of
(\ref{22}): presence of the background particles  will lead to
the following generating functional:
\be
R_{cp}=e^{iN(\hat{\phi}_i\hat{\phi}_j)}R_0(\phi_{\pm}),
\l{44}
\ee
where $R_0 (\p_{\pm})$ is the generating functional for
vacuum case, see (\ref{35}). The operator
$N(\hat{\p}^*_{i}\hat{\p}_{j}), i,j=+,-,$
describes the external particles environment.

The operator $\hat\p^*_i(q)$ can be considered as the creation and
$\hat\p_i(q)$ as the annihilation operator, see definition (\ref{13}).
Correspondingly the product $\hat\p^*_i(q)\hat\p_j(q)$ acts as
the activity operator. So, in the expansion of
$N(\hat\p^*_i\hat\p_j)$ we can leave only the first nontrivial
term:
\be
N(\hat{\phi}^*_i\hat{\phi}_j)=
\int d\omega (q) \hat{\phi}^*_i (q) n_{ij} \hat{\phi}_j (q),
\l{45}
\ee
since  no special correlation among background particles should be expected.
If the external (nondynamical) correlations are present then the
higher powers of $\hat\p^*_i\hat\p_j$ will appear in expansion (\ref{45}).
Following to the interpretation of $\hat\p^*_i\hat\p_j$ we conclude that
$n_{ij}$ is the mean multiplicity of background  particles.

In (\ref{45}) the  normalization condition:
\be
N(0)=0
\l{46}
\ee
was used and summation over all $i,j$ was assumed. In the vacuum
case only the combinations $i\neq j$ are present since the time ordering.

Computing $R_{cp}$ we   must  conserve the translational invariance of
amplitudes. It lead to extraction of the energy-momentum conservation
$\d$-functions. Having background particles flow it is important
to note that only connected contributions must be taken into account.
Using the Fourier transformation we can find that to each vertex of
in-going in $a_{n,m}$ particle we must adjust the factor
$e^{-i\a_1q/2}$ and for each out-going particle we have correspondingly
$e^{-i\a_2q/2}$.

So, the product $e^{-i\a_kq/2}e^{-i\a_jq/2}$ can
be interpreted as the probability factor of the one-particle
$(creation+annihilation)$ process. The $n$-particles
$(creation+annihilation)$ process' probability is the simple
product of  these factors if  there is not the special correlations
among background  particles. This interpretation is evident in the
CM frame $\a_k=(-i\b_k,\vec0)$.

After this preliminaries it is not too hard to find that in the
CM frame we have:
\ba
n_{++}(q_0)=n_{--}(q_0)=
\frac{\sum_{n=0}^{\infty}ne^{-\frac{\beta_1+\beta_2}{2}|q_0|n}}
{\sum_{n=0}^{\infty}e^{-\frac{\beta_1+\beta_2}{2}|q_0|n}}=
\n \\
\frac{1}{e^{\frac{\beta_1 +\beta_2}{2}|q_0|}-1}=
\tilde {n}(|q_0|\frac{\beta_1 +\beta_2}{2}).
\l{47}
\ea
Computing $n_{ij}$ for $i\neq j$ we must take into account that we have one
more particle:
\ba
n_{+-}(q_0)=
 \theta (q_0)
\frac{\sum_{n=1}^{\infty}ne^{-\frac{\beta_1+\beta_1}{2}q_0 n}}
{\sum_{n=1}^{\infty}e^{-\frac{\beta_1+\beta_1}{2}q_0 n}}+
 \Theta (-q_0)
\frac{\sum_{n=0}^{\infty}ne^{\frac{\beta_1+\beta_1}{2}q_0 n}}
{\sum_{n=0}^{\infty}e^{\frac{\beta_1+\beta_1}{2}q_0 n}}=
\n \\
 \Theta (q_0)(1+\tilde {n}(q_0 \beta_1))+
 \Theta (-q_0)\tilde {n}(-q_0 \beta_1)
\l{48}
\ea
and
\be
n_{-+}(q_0)=
 \Theta (q_0)\tilde {n}(q_0 \beta_2)+
 \Theta (-q_0)(1+ \tilde {n}(-q_0 \beta_2)).
\l{49}
\ee
Using (\ref{47}), (\ref{48}) and (\ref{49}), and
the definition (\ref{40}) we find the Green functions:
\be
G_{i,j}(x-x',(\beta))=\int \frac{d^4 q}{(2\pi)^4} e^{iq(x-x')}
\tilde{G}_ij (q, (\beta))
\l{50}
\ee
where
\ba
i\tilde{G}_ij (q, (\beta))=
\left( \matrix{
\frac{i}{q^2 -m^2 +i\epsilon} & 0 \cr
0 & -\frac{i}{q^2 -m^2 -i\epsilon} \cr
}\right)
+\n \\ \n \\+
2\pi \delta (q^2 -m^2 )
\left( \matrix{
\tilde{n}(\frac{\beta_1 +\beta_2}{2}|q_0 |) &
\tilde{n}(\beta_2 |q_0 |)a_+ (\beta_2) \cr
\tilde{n}(\beta_1 |q_0 |)a_- (\beta_1) &
\tilde{n}(\frac{\beta_1 +\beta_2}{2}|q_0 |) \cr
}\right)
\l{51}
\ea
and
\be
a_{\pm}(\beta)=-e^{\frac{\beta}{2}(|q_0|\pm q_0)}.
\l{52}
\ee
The corresponding
generating functional has the standard form:
\ba
R_{cp}(j_{\pm})=\exp\{-iV(-i\hat{j}_+)+iV(-i\hat{j}_-)\}
\times\n \\ \times
\exp\{\frac{i}{2}\int dx dx' j_i (x)G_{ij}(x-x',(\beta))j_j(x')\}
\l{53}
\ea
where the summation over repeated indexes  is assumed.

Inserting (\ref {53}) in the equation of state (\ref{33}) we can find that
$\beta_1 =\beta_2 =\beta (E)$. If $\beta (E)$ is a  ``good" parameter then
$G_{ij}(x-x';\beta )$ coincide with the Green functions of
the real-time finite-temperature field theory and the KMS boundary
condition:
\be
G_{+-}(t-t')=G_{-+}(t-t'-i\beta),\;\;\;
G_{-+}(t-t')=G_{+-}(t-t'+i\beta),
\l{54}
\ee
is restored. The eq.(\ref{54}) can be deduced from (\ref{51}) by
the direct calculations.

\section{Conclusion}
\setcounter{equation}{0}

In our interpretation of the  real-time  finite-temperature field theory
the statistics and fields quantum dynamics were unlinked:
statistics is fixed by the operator $\exp\{iN(\hat{\phi}^*_i \hat{\phi}_j\}$
and  a pure field-theoretical dynamics  is described by
$R_0 (\phi_{\pm})=Z(\phi_+ )Z^*(\phi_-)$, where $Z(\phi_{\pm})$
is the vacuum into vacuum transition amplitude in presence of the
external (auxiliary) fields $<vac|vac>_{\phi}$. We can say that the operator
$\exp\{iN(\hat{\p}_i\hat{\p}_j)\}$ maps the
system of interacting fields on the state with definite thermodinamical
parameters.

This means the following procedure: we can act by the operator exponent
$\exp \{ N \}$ before calculation of $R_0$. We find in this case from
(\ref{53}) that
\be
R_{cp}(\beta_1 ,\beta_2 )=\int D\Phi_+ D\Phi_-
e^{i\tilde{S}(\Phi_+)-i\tilde{S}(\Phi_-)}
e^{i\Psi (\beta;\Phi_{\pm})},
\l{5.1}
\ee
where $\tilde{S}(\Phi)$ is the total action and
\ba
\Psi (\beta;\Phi_{\pm})=-i \ln \{
e^{iV(\Phi_+ +\phi_+)- iV(\Phi_- +\phi_-)}
\n \\
e^{i\int dY dy[\hat{\phi}_i (Y+y/2)G_{ij}(y,(\beta  ))\hat{\phi}_j (Y-y/2)}
e^{-iV(\Phi_+ +\phi_+)+ iV(\Phi_- +\phi_-)} \}
\l{5.2}
\ea
is act as the effective source of fields $\Phi_{\pm}$ and depends from the
temperature $1/\beta_i$. Such representation introduce in
quantization procedure the external conditions of the considered
problem. This can help to calculate the observables with high accuracy.

The formalism allows to describe an arbitrary system. The special
interest presents the ``local equilibrium" case. The structure of $R_{cp}$
and  of Green functions remains the same for this case, see (\ref{53})
and (\ref{51}), but $\b_k$ becomes coordinate dependent: $\b_k=\b_k
(Y)$.  Note that $Y=(x+x')/2$ is the Wigner's coordinate
\C {carr,hu}. The derivation of this result will be given
in the following paper.

\vspace{0.2in}

{\Large \bf Acknowledgment}

\vspace{0.2in}

I would  like to thank my colleagues from the Inst. of Phys.
(Tbilisi), Inst. of  Math. (Tbilisi), Inst. of Nucl. Phys.
(St.-Petersburg), Joint Inst. of Nucl. Res. (Dubna), Inst. of Exp.
and Theor. Phys. (Moscow), Inst. of Theor. Phys. (Kiev) and
DAMPT (Cambridg, England), and especially I.Paziashvili, for
fruitful discussions which gave me the chance to extracting
the questions that were necessary to show up in this paper. The work
was supported in part by the  U.S. National Science Foundation.

\end{document}